\documentclass[journal]{IEEEtran}
\usepackage{graphicx}
\usepackage[caption=false,font=normalsize,labelfont=sf,textfont=sf]{subfig}
\usepackage{amsmath}
\usepackage{multirow}
\graphicspath{{figures/}}
\usepackage{multirow}

\begin{document}
\title{Single-View 3D Reconstruction of Correlated Gamma-Neutron Sources}

%

\author{Mateusz~Monterial,~\IEEEmembership{Student Member,~IEEE},
        Peter~Marleau,~\IEEEmembership{Member,~IEEE},
        and~Sara~A.~Pozzi,~\IEEEmembership{Senior Member,~IEEE}
\thanks{This material is based upon work supported by the U.S. Department of Homeland Security Grant Award Number: 2012-DN-130-NF0001. The views and conclusions contained in this document are those of the authors and should not be interpreted as representing the official policies, either expressed or implied, of the U.S. Department of Homeland Security.}
\thanks{Sandia National Laboratories is multi-program laboratory managed and operated by Sandia Corporation, a wholly owned subsidiary of Lockheed Martin Corporation, for the U.S. Department of Energy's National Nuclear Security Administration under contract DE-AC04-94AL85000. SAND number 2016-10226 J.}
\thanks{This work was funded in-part by the Consortium for Verification Technology under Department of Energy National Nuclear Security Administration award number DE-NA0002534.}
\thanks{M. Monterial and S. A. Pozzi are with the Nuclear Engineering and Radiological Sciences Department, University of Michigan, Ann Arbor, MI 48109 USA (e-mail: mateuszm@umich.edu).}%
\thanks{P. Marleau is with Sandia National Laboratories, Livermore, CA 94550 USA.}%
}

\maketitle
\pagestyle{empty}
\thispagestyle{empty}

\begin{abstract}
We describe a new method of 3D image reconstruction of neutron sources that emit correlated gammas (e.g. Cf-252, Am-Be). This category includes a vast majority of neutron sources important in nuclear threat search, safeguards and non-proliferation. Rather than requiring multiple views of the source this technique relies on the source's intrinsic property of coincidence gamma and neutron emission. As a result only a single-view measurement of the source is required to perform the 3D reconstruction. In principle, any scatter camera sensitive to gammas and neutrons with adequate timing and interaction location resolution can perform this reconstruction. Using a neutron double scatter technique, we can calculate a conical surface of possible source locations. By including the time to a correlated gamma we further constrain the source location in three-dimensions by solving for the source-to-detector distance along the surface of said cone. As a proof of concept we applied these reconstruction techniques on measurements taken with the the Mobile Imager of Neutrons for Emergency Responders (MINER). Two Cf-252 sources measured at 50 and 60 cm from the center of the detector were resolved in their varying depth with average radial distance relative resolution of 26\%. To demonstrate the technique's potential with an optimized system we simulated the measurement in MCNPX-PoliMi assuming timing resolution of 200 ps (from 2 ns in the current system) and source interaction location resolution of 5 mm (from 3 cm). These simulated improvements in scatter camera performance resulted in radial distance relative resolution decreasing to an average of 11\%.
\end{abstract}


\section{Introduction}

\IEEEPARstart{T}{he} practice of radiation imaging is well established in fields as diverse as medicine \cite{Wernick2004, Todd1974}, astronomy \cite{Schönfelder2004, Herzo1975}, and nuclear safeguards and non-proliferation \cite{Phillips1995, Phillips1997}. Radiation imaging cameras function by either: 1) modulating the incident radiation, or 2) tracking the multiple scatters of incident particle in the detector medium. Hal Anger was the first to develop a gamma camera through the use of multichannel collimators to modulate incident radiation \cite{Anger1964}. The same principle can be applied with more complex coded aperture masks, which are analogous to superimposed pinhole cameras, to image both thermal and fast neutrons \cite{Vanier2004, Marleau2010}. 

The technique introduced in this paper is an extension of the second category of radiation cameras which track multiple scatters to reconstruct source location. In gamma ray imaging, these are called Compton cameras \cite{Phillips1995}, and are a mature technology with commercially available portable cameras \cite{Wahl2015}. The neutron scatter camera functions in an analogous way to the Compton camera, but with the use of the time-of-flight (TOF) between the first two scatters to determine the incident neutron energy \cite{Legge1968}. 

The discussion so far has been limited to 2D imaging systems, but in principle any radiation camera can produce 3D reconstruction of a source. The most common approach is to take multiple 2D images from different views, and combine them to form 3D rendering of the source. This technique is used in Single Photon Emission Computed Tomography (SPECT) and Positron Emission Tomography (PET) to image radioisotopes inside a patient \cite{Wernick2004}. More recently researchers at Lawrence Berkeley National Laboratory have used a variation of this technique, combined with 3D rendering of physical space, to reconstruct source locations in real time \cite{Barnowski2015}. All of these techniques require multiple views of the source and some freedom of movement. 

Single-sided 3D imaging has been demonstrated in Compton cameras by taking advantage of the parallax effect \cite{McKisson1994}. However, parallax techniques require a large solid angle coverage to function as a modality at all, whereas our method only requests solid angle coverage to increase efficiency. Furthermore, our technique would function at any distance with a portable system, even if it required long dwell times. Parallax by comparison would be restricted by system size and require close enough source-to-detector distances to function in the first place. This makes our technique potentially valuable for nuclear inspection, emergency response and treaty verification, where multiple views of the object of interest may be restricted, even though the location of the source is known. 

A single system can image both neutrons and gammas if the underlying detectors are sensitive to both particle types. Such scatter camera systems were constructed as radiation telescopes for astronomy \cite{Herzo1975}, and more recently for stand-off detection in nuclear security applications \cite{Poitrasson2014}. We employ this class of scatter camera to project a cone of possible source locations using the double neutron scatter principle. Source location is then further bound to a slice of the cone by adding a further constraint imposed by the time to a correlated gamma. Therefore each detected neutron double scatter with a coincident gamma yields a ring of possible source locations in space. In this regard our technique is similar to PET which requires the simultaneous emission of two gammas, detection of which yields a line of possible source locations. This new imaging technique utilizes an inherent property of majority of neutron sources important in nuclear threat-search, safeguards and non-proliferation: the coincident emission of neutrons and gammas. These sources include those undergoing spontaneous and induced fission \cite{Gozani2009}, a property of Special Nuclear Materials (SNM), and common ($\alpha$,n) sources that leave the remaining nucleus in an excited state leading to prompt gamma emission \cite{Geiger1975}. From the point of view of our detectors the emission of prompt gammas is essentially simultaneous to the emission of coincident neutrons. 

\section{Methods}

\subsection{Neutron double scatter imaging}

Measurement of a double-scatter neutron provides a conical surface of possible source locations with the vertex at the first neutron scatter ($n_0$) and axis defined by the location of two scatters. In a traditional segmented scatter camera, each scatter is measured as a separate interaction within any two detector cells of the measurement system. The incident neutron energy and by extension velocity of the incident neutron ($v_n$) can also be calculated from this measurement \cite{Legge1968}. 

The outgoing energy following the first neutron scatter is calculated by the time-of-flight ($\Delta t_{n_0,n_1}$) to the second scatter:
\begin{align} \label{eq:n_tof}
E_{n1} &= \frac{m_n}{2}\left(\frac{d_n}{\Delta t_{n_0,n_1}} \right)^2 
\end{align}
where $d_n$ is the distance between the two scatters. The outgoing energy is then summed with the energy lost due to proton recoil ($E_p$) in the first scatter which gives the initial incident energy of the neutron:

\begin{align} \label{neutron_erg}
E_{n0} &= E_p + E_{n1}.
\end{align}

\begin{figure}[!h]
	\centering
	\includegraphics[width=1.7in]{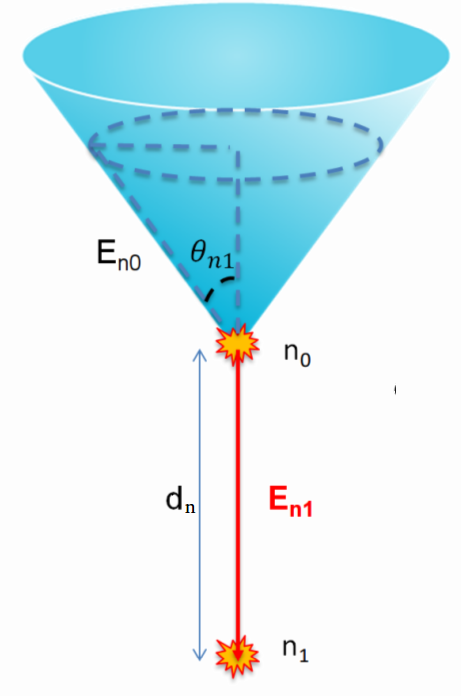}
	\caption{Illustration of the kinematics of a double neutron scatter in a scatter camera, resulting in a cone of possible source locations.} 
	\label{fig:neutron_double}
\end{figure}

From this we can establish the opening angle of the cone of possible source locations by
\begin{align} \label{eq:cone}
\cos^2(\theta_{n1}) = \frac{E_{n1}}{E_{n0}}.
\end{align} 
The cone has its axis defined by a vector from the second to first scatter. The resulting cone projection of source locations is illustrated in Fig. \ref{fig:neutron_double}. The next step for 2D neutron imaging is to chose a reasonable projection distance, and display the image formed from the overlapping regions of the projected cones. We will demonstrate that by measuring a coincident gamma with a double scattered neutron it is possible to calculate the distance from the first neutron scatter to the possible source locations along the surface of the cone. 

\subsection{New method for 3D reconstruction}

\begin{figure}[!h]
	\centering
	\includegraphics[width=3.5in]{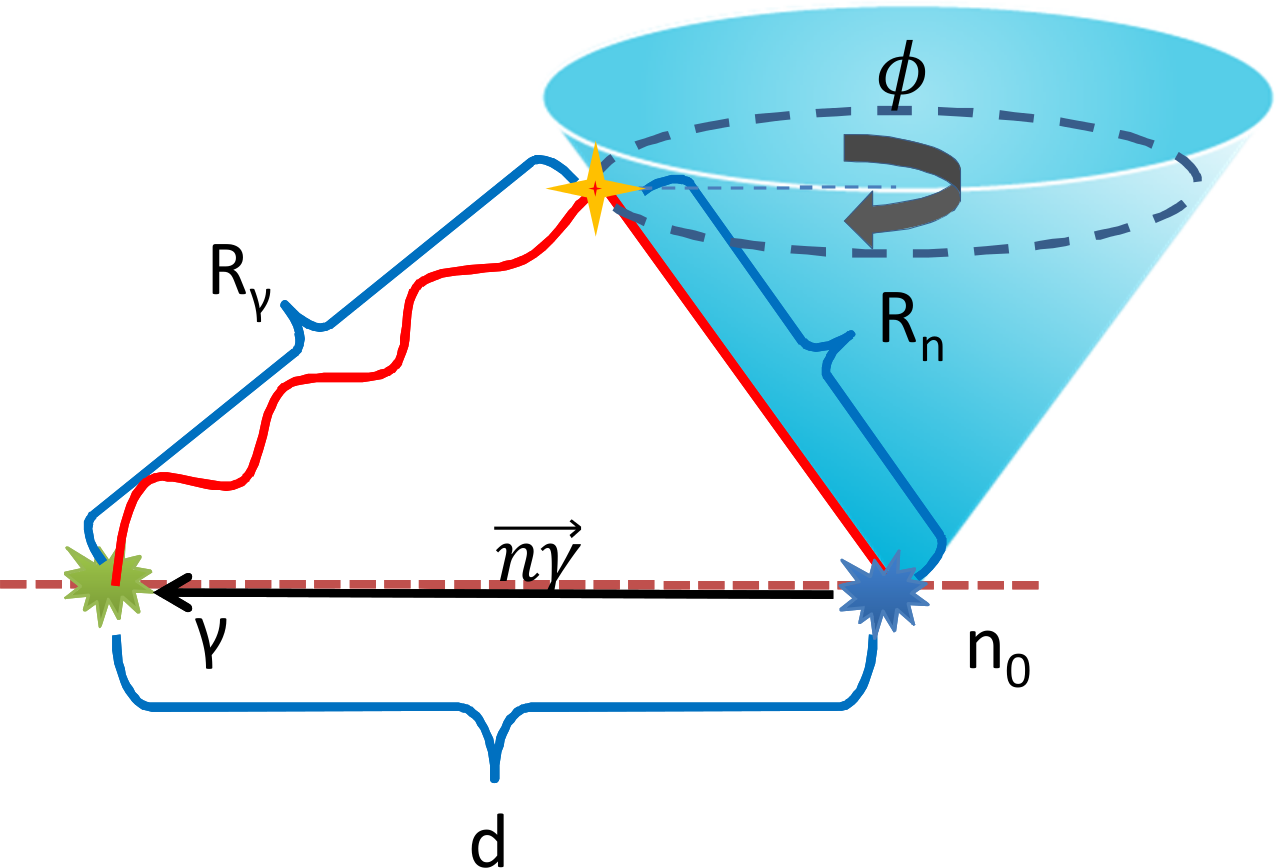}
	\caption{The cone of possible source locations from neutron double scatter and a corresponding correlated gamma. The second neutron scatter is not shown. The distances between the source (yellow 4-pointed star) and first neutron scatter ($R_n$) and gamma ($R_\gamma$) are shown for one of the possible locations along the surface of the cone. All other possible source locations lay somewhere along the azimuthal ($phi$) angle of the cone.}
	\label{fig:cone}
\end{figure}

Our new imaging technique requires the detection of gammas ($\gamma$) in coincidence with double scattered neutrons. This is made possible with fast organic scintillators capable of discriminating between incident neutrons and gammas. An illustration of the first neutron scatter ($n_0$) and the correlated gamma is shown in Fig. \ref{fig:cone}. $R_n$ and $R_\gamma$ are the distances from a possible source location to the neutron and gamma interactions, respectively. The goal is to solve for $R_n$ in order to constrain the possible source location to the third dimension along the surface of the cone.

The two unknown distances $R_n$ and $R_\gamma$, and the known distance $d$ triangulate the location of the source along the azimuthal angle ($\phi$) around the cone of possible source locations. We use the law of cosines to relate these variables in a parametric equation
\begin{align} \label{eq:cosine}
R_\gamma^2 = R_n^2 + d^2 - 2 R_n \mu d
\end{align}
where $\mu$ is the cosine of the angle between the cone surface and the vector $\overrightarrow{n \gamma}$
\begin{align} \label{eq:mu}
\mu = \frac{\overrightarrow{n \gamma}}{d} \cdot \hat{R_n}(\phi).
\end{align}
$\hat{R_n}$ is a unit vector pointing from the vertex of the cone to any possible source location along cone surface. 

Along with implicitly measuring the locations of $\gamma$ and $n_0$ interactions we also measure the time difference ($\Delta t$) between them. This gives us the second equation that relates $R_n$ and $R_\gamma$:
\begin{align} \label{eq:delT}
\Delta t = \frac{R_n}{v_n} - \frac{R_\gamma}{c}.
\end{align}

Using (\ref{eq:cosine}) and (\ref{eq:delT}) we solve the quadratic equation for the distance $R_n$ from the first neutron scatter to the source. Given that the velocity of the neutron is always less than the speed of light, $v_n < c$, the only valid solution for this distance is
\begin{align} \label{eq:Rn_final}
R_n = & (c^2 - v_n^2)^{-1} \Big[  c^2 \Delta t v_n - dv_n^2 \mu  \\ 
 & + \sqrt{ v_n^2\left(c^2(\Delta t^2v_n^2 -2dv_n\mu \Delta t +d^2) + v_n^2d^2(\mu^2-1) \right) } \Big]. \notag
\end{align}

\begin{figure}[!h]
	\centering
	\includegraphics[width=3.5in]{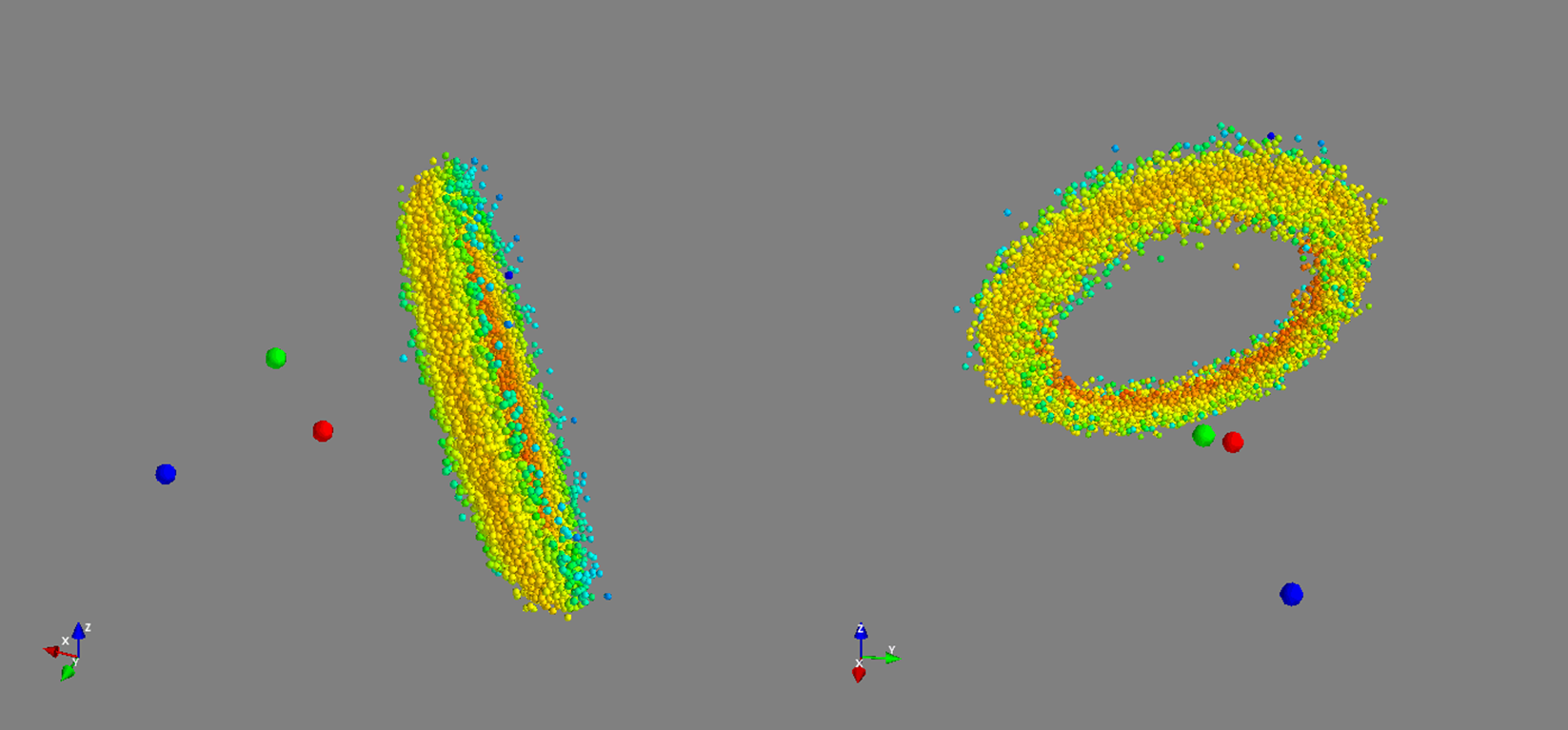}
	\caption{Possible source locations for a single measured correlated events shown as colored spheres. The first (red) and second (blue) neutron scatter define the central axis of the cone and the opening angle, and the correlated gamma (green) constrains the radial distance to form the resulting ``donut" shape. The superposition of many donuts will reveal the source location in the overlapping region. For illustrative purposes we show the same object from two different angles.}
	\label{fig:donut}
\end{figure}

The parametric solution to (\ref{eq:Rn_final}) allows us to define a slice of possible source locations from the double neutron scatter cone. Therefore, measurement of a double-scattered neutron with a correlated gamma will yield a ``donut" in 3D space, as shown in Fig. \ref{fig:donut}, of possible source locations. This torus-like shape is analogous to PET's line-of-response for a measured pair coincident gammas. Multiple such events further constrain the distribution of possible source locations by superposition of the torus-like shapes in 3D space. The region of overlap among those shapes will reveal the true source location.  

\subsection{Stochastic Origin Ensemble}

Stochastic Origin Ensemble (SOE) was used in an effort to improve the final reconstructed image. It is an application of the Metropolis-Hastings algorithm which improves the reconstruction quality over standard back-projection. It is not required for image reconstruction in our technique, but it was used here because it has been shown to improve image quality of dual-particle imaging systems \cite{Hamel2016}. 

The basic idea behind SOE is to sample the measured quantities (time, interaction location, energy resolution) with appropriate uncertainties and estimate the source distribution as a probability density function (PDF). The PDF is estimated as the density of source locations averaged over all iterations of the SOE algorithm. Details of the SOE algorithm as applied to Compton imagers is given in \cite{Andreyev2011}. 

The rarity of measuring a coincident gamma with a double scattered neutron meant we only had a few thousand events to perform 3D reconstruction. Accurate estimation of the source PDF, a key step between iterations in SOE, is made difficult by the sparsity of the 3D space. As a solution a Kernel Density Estimator (KDE) was used to calculate the density at each source location. We found that the Epanechinkov kernel provided optimal results at reasonable computing times. The bandwidth parameters that were close to the resolution of the system (2-6 cm) appeared to work the best. The Scikit-learn machine learning package for Python was used to compute the requisite density estimations \cite{scikit-learn}.

\section{Experiments}

To test this technique we used the Mobile Imager of Neutrons for Emergency Responders (MINER), developed by Sandia \cite{Goldsmith2014}. It is an array of sixteen 7.62$\times$7.62 cm EJ-309 cylindrical detectors packaged in a larger cylindrical form-factor for portability and symmetry which allows for omnidirectional (4$\pi$) imaging. The portability comes at the expense of imaging resolution because adjacent cells centers are only 11.9 cm apart.

\begin{figure}[!h]
	\centering
	\includegraphics[width=3in]{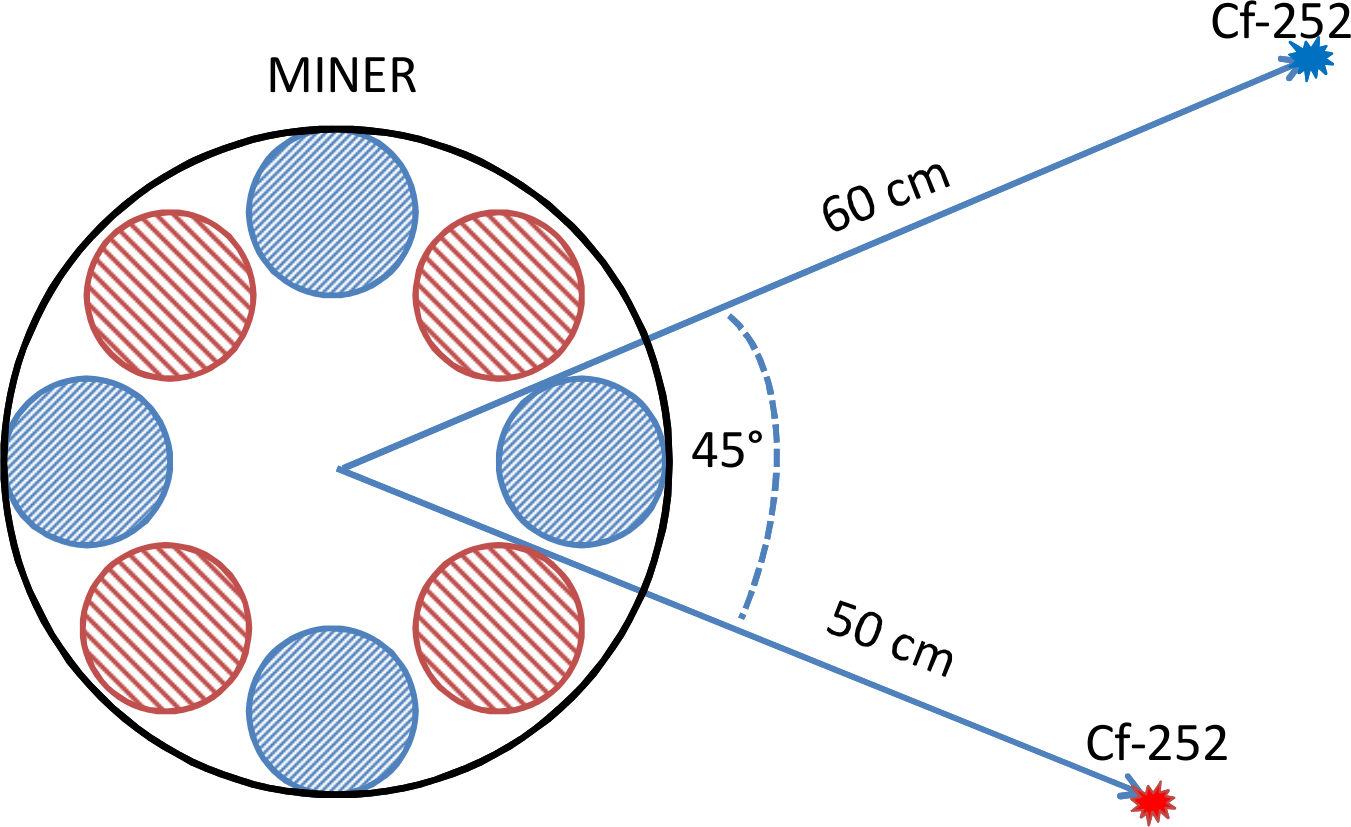}
	\caption{Measurement configuration showing the position of the two Cf-252 sources with respect to MINER. The dimensions of the detector and source-to-detector distances are drawn in correct proportions.}
	\label{fig:setup}
\end{figure}

Two equal strength, 26.7 $\mu$Ci, Cf-252 sources were measured 50 and 60 cm away from the center of MINER. The two sources were placed 45$^{\circ}$ apart as shown in Fig. \ref{fig:setup}. The centers of each detector cell were taken as the position of the incident particle interaction. MINER has a timing resolution of about 2 ns and interaction location resolution of at least 3 cm. 

To demonstrate the 3D reconstructions best achievable capability we also simulated the same experiment, with MINER, and assumed timing resolution of 200 ps and interaction location resolution of 5 mm. The former is possible with fast photomultiplier tubes (PMTs) or silicon photomultipliers (SiPMs) \cite{Cattaneo2016}. The latter can be achieved by using smaller detector cells, or using multiple readouts to better localize the event of interaction. MCNPX-PoliMi was used to perform the requisite simulations \cite{Pozzi2012}.

\section{Results and Discussion}

\subsection{Image Reconstruction}

\begin{figure}[!t]
	\centering
	\includegraphics[width=3in]{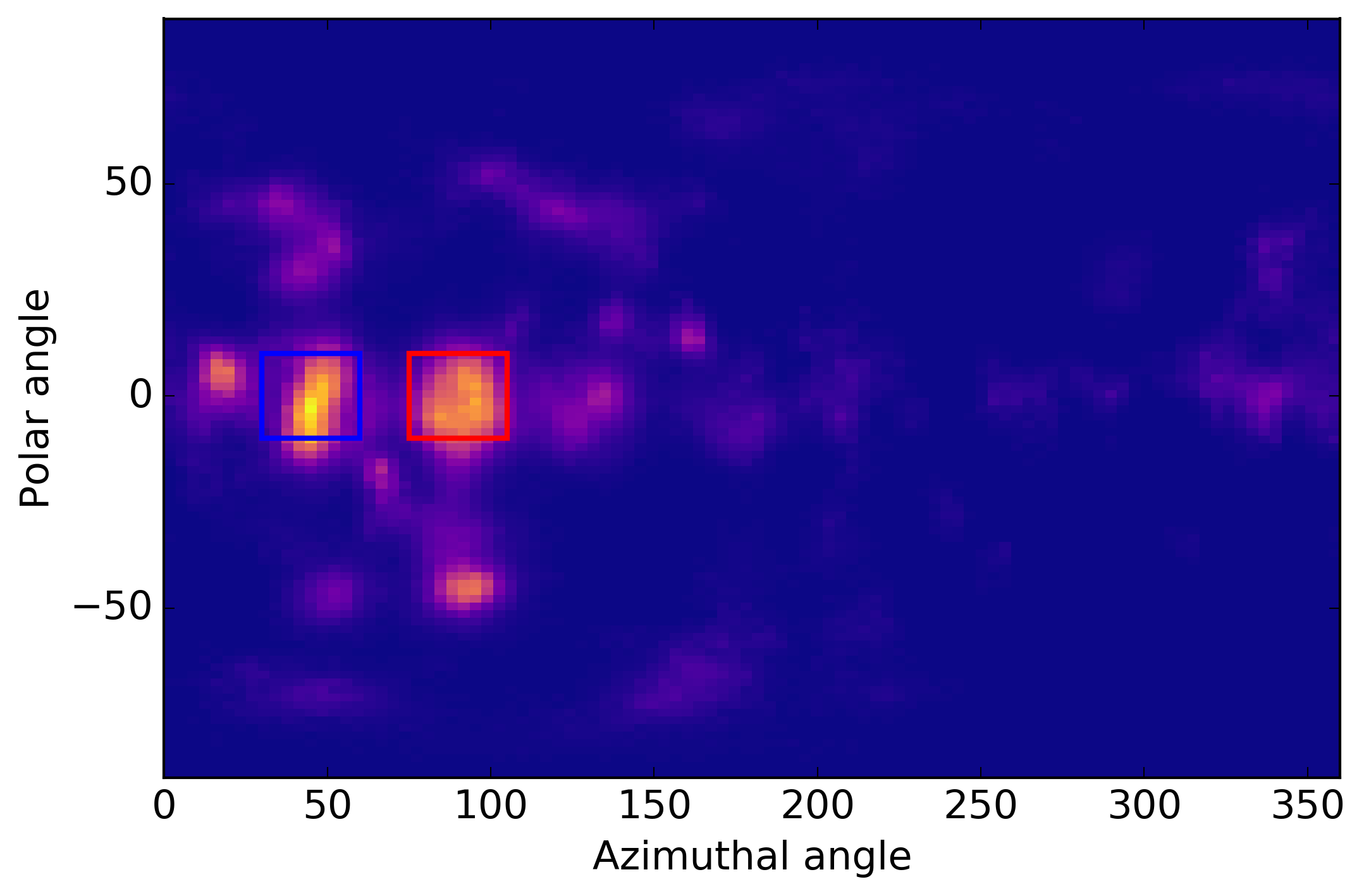}
	\includegraphics[width=3in]{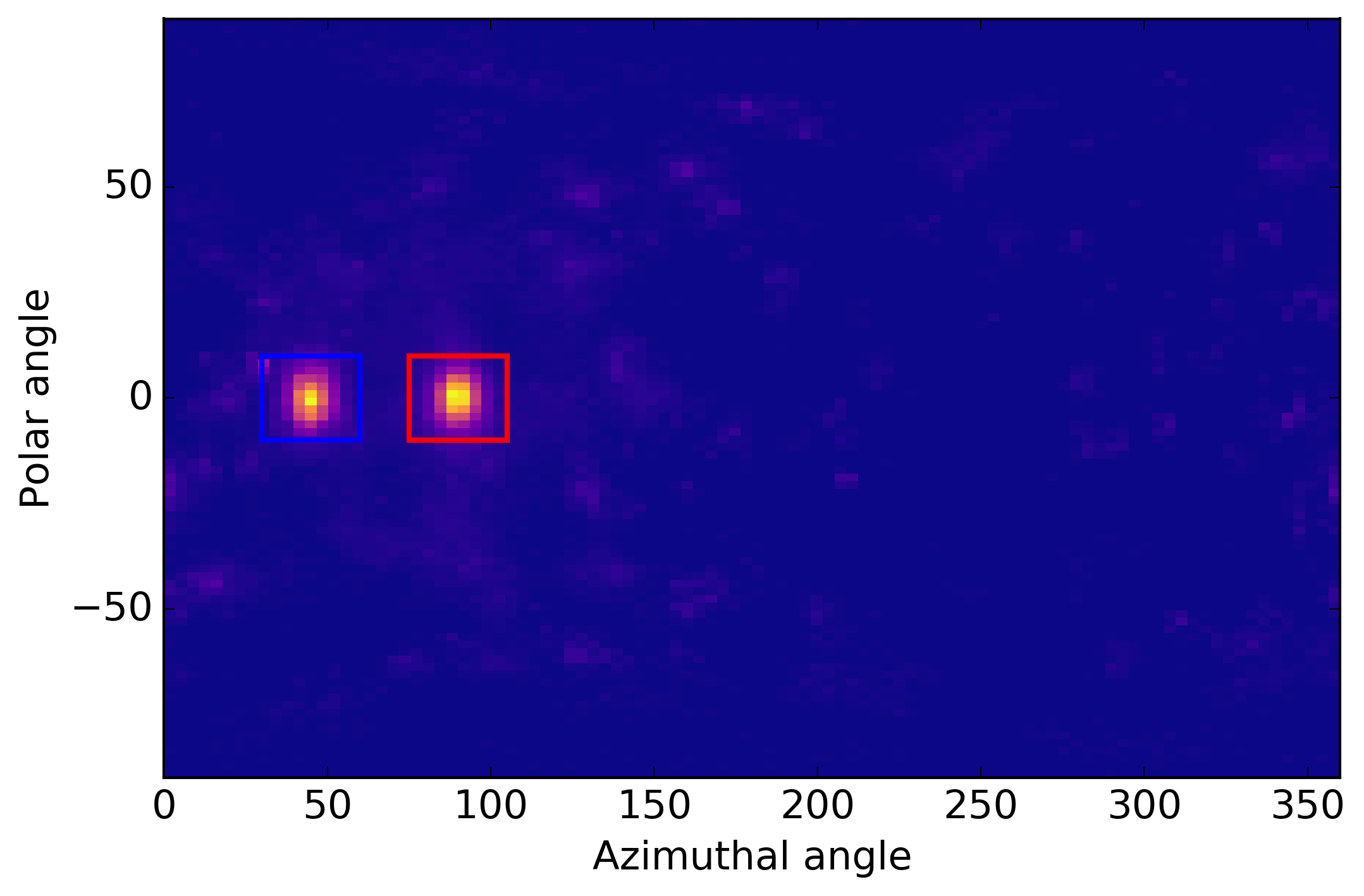}
	\caption{The measurement (\textit{top}) and simulated (\textit{bottom}) images with each reconstructed source point weighted by $r^4$.  Each source is marked by a blue (60 cm source) and red (50 cm source) square.}
	\label{fig:side_view}
\end{figure}

Admittedly it is a challenge to display 3D results on a two-dimensional sheet of paper. The reconstruction method used provides the Cartesian coordinates $(x,y,z)$ of possible source locations. We converted these to spherical coordinates and chose to display the bi-variate histograms of the polar and azimuthal angles in Fig. \ref{fig:side_view}. This bi-variate histogram is essentially a 2D image of the source from the point of view of the detector. We marked each source by red (50 cm source) and blue (60 cm source) squares. The source points from those marked regions were used to estimate the radial distance and azimuthal angle resolutions.

\begin{figure}[!t]
	\centering
	\includegraphics[width=3in]{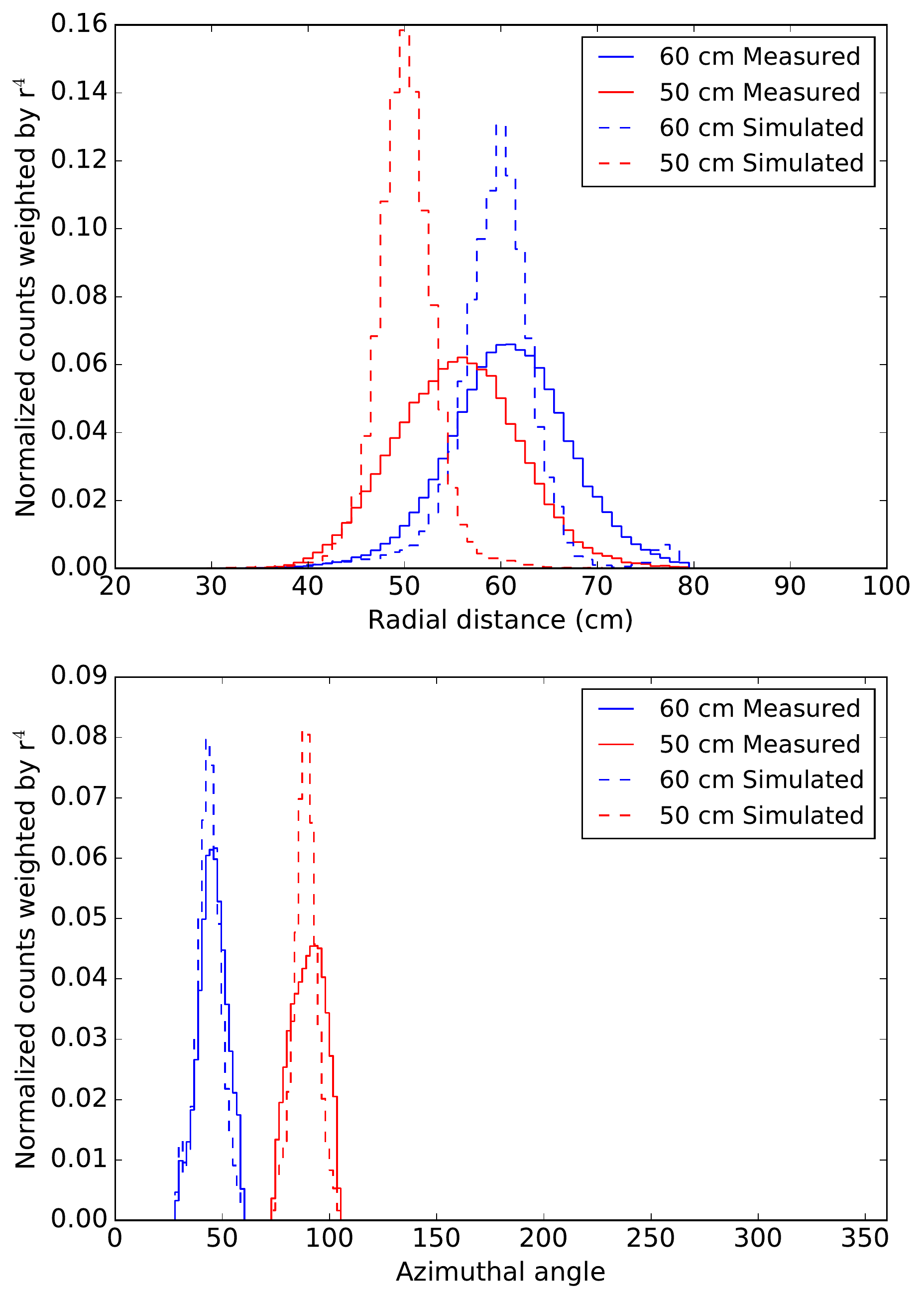}
	\caption{The radial distance (\textit{top}) and azimuthal angle (\textit{bottom}) distributions for both measurement (\textit{solid}) and simulation (\textit{dashed}). The radial distance distribution describes the distance from detector center. The source points were taken from the within the squares of the images in Figure \ref{fig:side_view}, with matching color combinations.}
	\label{fig:theta_r_all}
\end{figure}

\begin{figure}[h]
	\centering
	\includegraphics[width=3in]{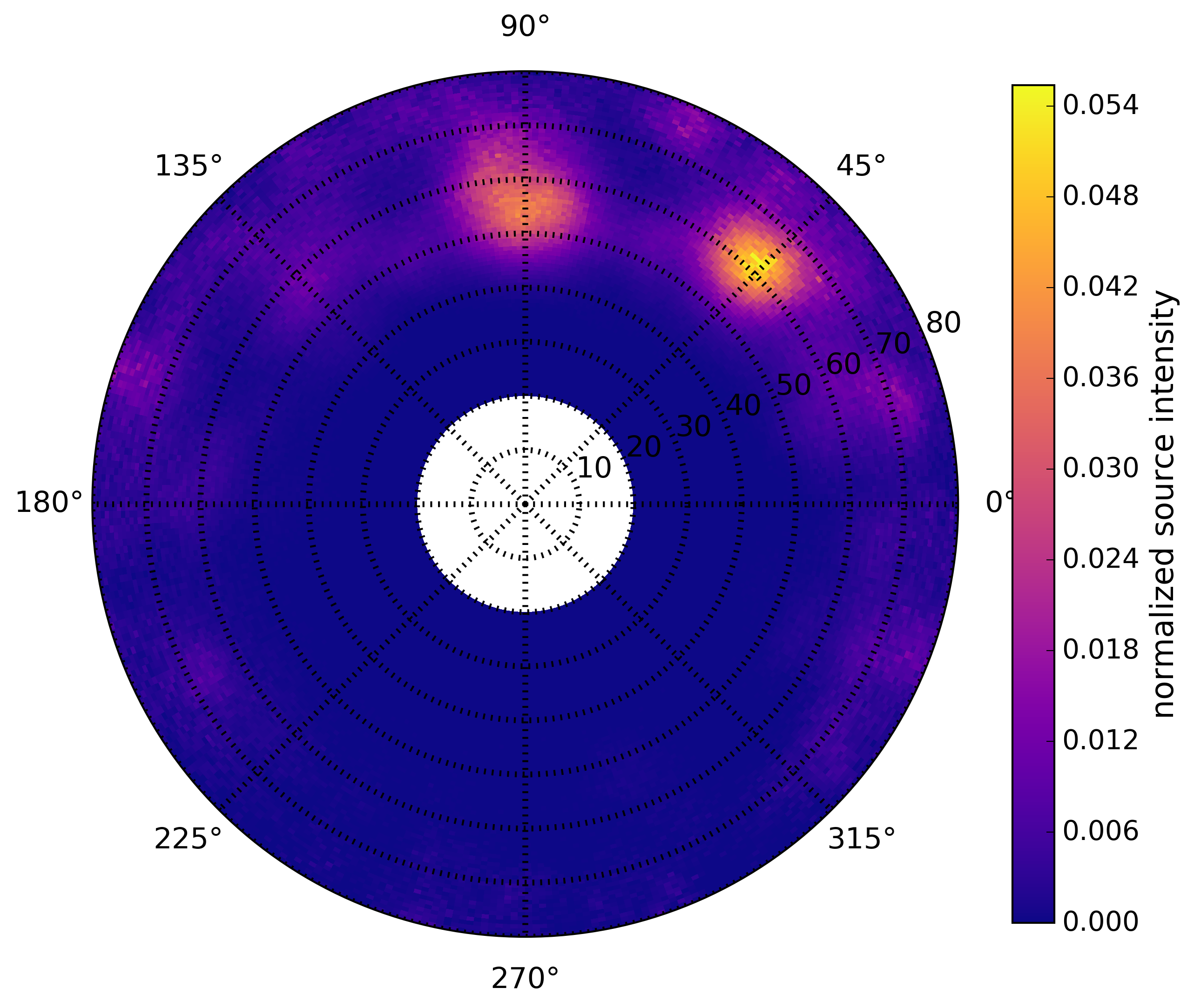}
  \includegraphics[width=3in]{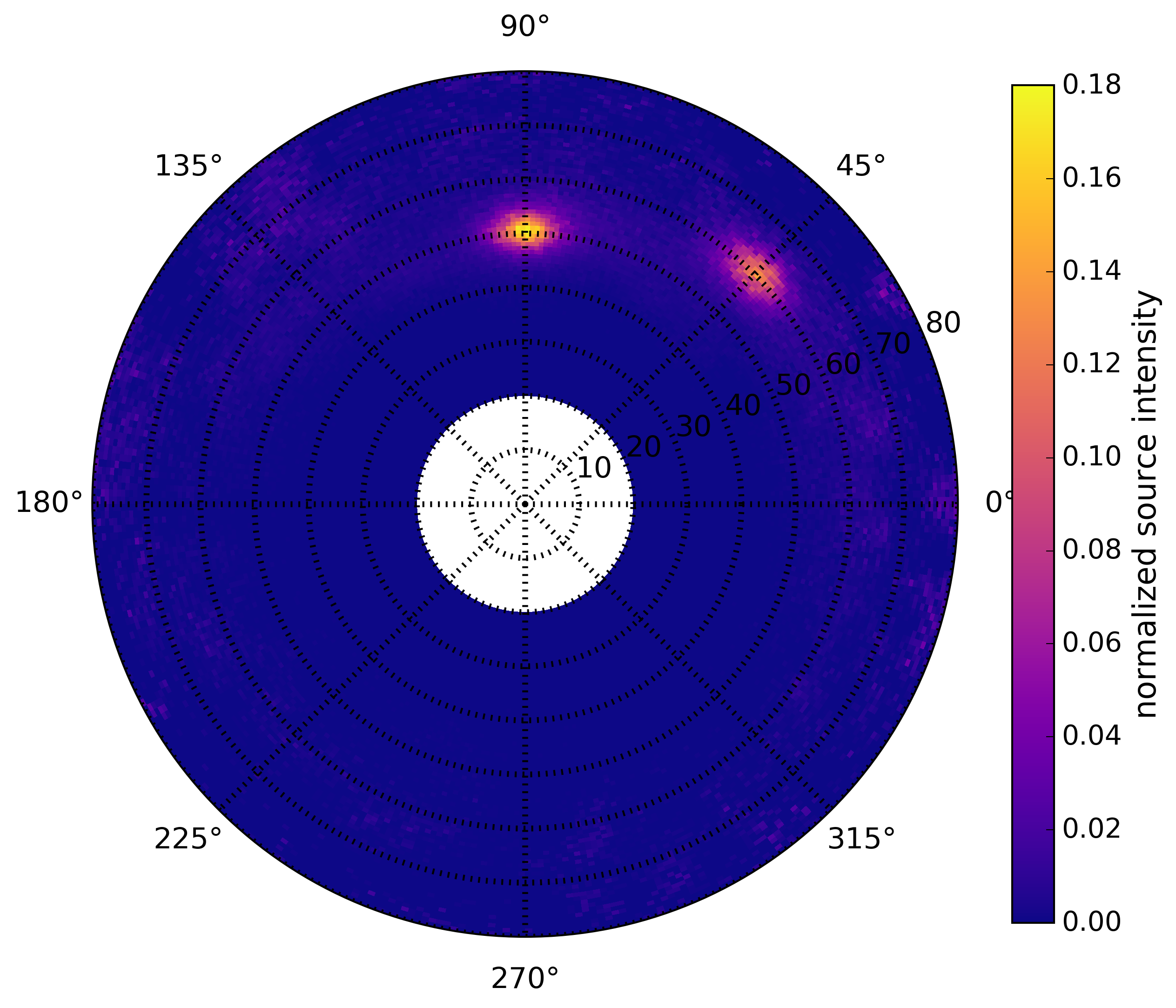}
	\caption{The polar projection (top-down view) of the image reconstruction for both the (\textit{top}) measurement  and (\textit{bottom}) simulation.}
	\label{fig:top_view}
\end{figure}

The corresponding histograms of the radial distance ($r$) to the detector and the azimuthal angle for each source are shown in Fig. \ref{fig:theta_r_all}. Fig. \ref{fig:top_view} shows a top-down view of the sources by looking at source particle densities projected on a polar plane. In all images and distributions shown each source point was weighted by $r^4$ in order to account for the efficiency of detecting two correlated particles. 

In both measurement and simulation the two sources are clearly resolved in both radial distance and angular space. The 2D image from the measurement shows some reconstruction artifacts, likely caused by the application of SOE. These are largely due to the choice of a narrow bandwidth parameter, which improves resolution at the cost of having these artifacts in the final reconstruction. Nevertheless, the final image reconstruction of these point sources was not very sensitive to the selection of a bandwidth parameter from 2 to 10 cm. 

\subsection{Radial Distance and Angular Resolutions}

We estimated the radial distance and angular resolutions for both measurement and simulation from the isolated source point distributions shown in Fig. \ref{fig:theta_r_all}. In order to estimate resolution we looked at the Full Width Half-Maximums (FWHM) of each distribution and also the peak locations (or centers) for a measure of accuracy. Both parameters were estimated by interpolating through the points in the distributions. The radial distribution and angular distribution parameters are shown in Table \ref{tab:radial} and \ref{tab:azimuthal}, respectively. 

The radial distance relative resolution, the ratio of FWHM and peak location, averaged at 26\% for the measurement and 11\% for the simulation. The angular resolution for the measurement improved from FWHM of 23$^{\circ}$ to 15$^{\circ}$ with source-to-detector distance. The simulation results have an improved azimuthal FWHM of 11$^{\circ}$ for both sources. 

There is an apparent skewness of the radial distance distribution of the 60 cm source in the direction of the 50 cm source, which may contribute to the increase in absolute FWHM. This is due to some 50 cm source points present in the same angular region of the 60 cm source. It is purely a matter of the nearly double relative efficiency of detecting correlated signature from a source that is 20\% closer. 

The peak location for the measurement is off by 5.8 cm for the 50 cm source. If the source locations are not weighted by radial distance this discrepancy drops to 2.1 cm. By contrast the 60 cm source peak location is off by only 1.5 cm. This shift in peak location is not exhibited in the simulated system with better timing and interaction location resolution. We believe that this effect is caused by both the weighting of source locations and the presence of the 60 cm source. Weighting source location by radial distance taken to the fourth power is necessary in order to account for the $r^{-4}$ drop in a efficiency. However, the weighting skewed the radial distribution of the 50 cm source toward the 60 cm source due to the significant overlap in the both radial distribution. By contrast the radial distributions of the simulated sources were well separated, as seen in Fig. \ref{fig:theta_r_all}, and the peak locations matched the true locations of the sources.

\begin{table}
\caption{Radial distribution (units in cm) parameters for each source in both measurement and simulation}
\centering
\begin{tabular}{l || c c c} 
  \hline \\ [-1.5ex]
  & Source Distance & Peak Location & FWHM \\ \hline \\ [-1.5ex]
\multirow{2}{*}{Measurement} & 50  & 55.8 & 13.5 \\
 & 60  & 61.5 & 15.1 \\ \hline \\ [-1.5ex]
\multirow{2}{*}{Simulation} & 50  & 50.6 & 5.7 \\
 & 60  & 60.3 & 6.8 \\
\end{tabular}  \label{tab:radial}
\end{table}

\begin{table}
\caption{Azimuthal angular distribution parameters for each source in both measurement and simulation}
\centering
\begin{tabular}{l || c c c} 
  \hline \\ [-1.5ex]
  & Angular Position & Peak Location & FWHM \\ \hline \\ [-1.5ex]
\multirow{2}{*}{Measurement} & 45$^{\circ}$  & 47$^{\circ}$ & 15$^{\circ}$ \\
 & 90$^{\circ}$  & 91$^{\circ}$ & 23$^{\circ}$ \\ \hline \\ [-1.5ex]
\multirow{2}{*}{Simulation} & 45$^{\circ}$  & 45$^{\circ}$ & 11$^{\circ}$ \\
 & 90$^{\circ}$  & 90$^{\circ}$ & 11$^{\circ}$ \\
\end{tabular}  \label{tab:azimuthal}
\end{table}

\subsection{Uncertainty Quantification}

Uncertainty quantification was performed using linear error propagation theory \cite{Lebigot2010} to determine error contributions from timing and interaction location resolutions on the radial source-to-detector distance. The overall timing resolution had over twice the error contribution compared with the interaction location resolution. The gamma timing was nearly six times more important than the neutron timing, which makes sense given the relative speed of each particle. By contrast the neutron interaction location contributed nearly an order of magnitude more compared with the gamma interaction location. The first neutron interaction location had double the contribution of the second interaction. 

In conclusion, the gamma timing resolution and neutron interaction location resolution are the primary contributors to uncertainty in radial source-to-detector distance. This was true for the measurement with timing resolution of 2 ns and interaction location resolution of 3 cm. At the simulated resolutions of 200 ps and 5 mm, the uncertainty in the proton recoil energy from the first neutron scatter ($E_p$) was the limiting factor. This includes the effects of energy resolution and calibration of the detector and measurement of light output response. The current practice involves fitting light output response to an empirical formula \cite{Enqvist2013, Pino2014}. Improvements could be made by better characterization of the detector response, but the estimation of the proton recoil energy would ultimately be limited by the low energy resolution of organic scintillators. 

\section{Conclusions}

We developed a method for 3D reconstruction of sources that emit correlated gammas and neutrons. The technique is distinguished from traditional 3D radiation imaging methods by only requiring a single-sided measurement of the source. And unlike parallax, it only requests large solid angle coverage to increase efficiency not enable function. The method proposed here is an extension of double neutron scatter imaging, combined with a correlated gamma to constrain the source location to the third dimension. 

It is possible to resolve two sources of equal strength 10 cm apart using a portable scatter camera with sub-optimal timing and angular resolution. Simulated results show a four fold improvement in depth and angular resolution if the timing and interaction location resolution were improved. Such improvements could be attainable with current technology. Furthermore the detector cell geometry and size could be further optimized to improve localization resolution while maintaining adequate efficiency. 

The efficiency of detecting a correlated neutron-gamma pair decreases as $r^{-4}$, where $r$ is the radial distance between source and detector system center. Therefore it is not an ideal technique for standoff detection, although efficiency could be scaled with number and size of detector cells. However, this technique could prove valuable in application where access to the object of interest is limited. For example, this could include inspection of nuclear facilities for safeguards or treaty verification.Furthermore, neutron sources that emit correlated gammas (e.g. fission, ($\alpha$,n)) are ubiquitous and include the vast majority of sources of concern in the above mentioned applications.

\appendices


\section*{Acknowledgment}
The authors would like to thank John Goldsmith for lending MINER, and Michael Hamel for guidance on the SOE method. 


%
\bibliographystyle{ieeetr}
\bibliography{ieee2015}

\begin{thebibliography}{10}

\bibitem{Wernick2004}
M.~Wernick and J.~Aarsvold, {\em Emission Tomography: The Fundamentals of PET
  and SPECT}.
\newblock Elsevier Science, 2004.

\bibitem{Todd1974}
R.~W. Todd, J.~M. Nightingale, and D.~B. Everett, ``A proposed [gamma]
  camera,'' {\em Nature}, vol.~251, pp.~132--134, Sep 1974.

\bibitem{Schönfelder2004}
V.~Schönfelder, ``Imaging principles and techniques in space-borne gamma-ray
  astronomy,'' {\em Nuclear Instruments and Methods in Physics Research Section
  A: Accelerators, Spectrometers, Detectors and Associated Equipment},
  vol.~525, no.~1–2, pp.~98 -- 106, 2004.
\newblock Proceedings of the International Conference on Imaging Techniques in
  Subatomic Physics, Astrophysics, Medicine, Biology and Industry.

\bibitem{Herzo1975}
D.~Herzo, R.~Koga, W.~Millard, S.~Moon, J.~Ryan, R.~Wilson, A.~Zych, and
  R.~White, ``A large double scatter telescope for gamma rays and neutrons,''
  {\em Nuclear Instruments and Methods}, vol.~123, no.~3, pp.~583 -- 597, 1975.

\bibitem{Phillips1995}
G.~W. Phillips, ``Gamma-ray imaging with compton cameras,'' {\em Nuclear
  Instruments and Methods in Physics Research Section B: Beam Interactions with
  Materials and Atoms}, vol.~99, no.~1, pp.~674 -- 677, 1995.

\bibitem{Phillips1997}
G.~W. Phillips, ``Applications of compton imaging in nuclear waste
  characterization and treaty verification,'' in {\em Nuclear Science
  Symposium, IEEE}, (Albuquerque, NM), Nov 9-15 1997.

\bibitem{Anger1964}
H.~O. Anger, ``Scintillation camera with multichannel collimators,'' {\em J
  Nucl Med.}, vol.~5, pp.~583 -- 531, July 1964.

\bibitem{Vanier2004}
P.~Vanier, ``Improvements in coded aperture thermal neutron imaging,'' in {\em
  SPIE Conference Proceedings}, vol.~5199, p.~124, 2004.

\bibitem{Marleau2010}
P.~Marleau, J.~Brennan, E.~Brubaker, and J.~Steele, ``Results from the coded
  aperture neutron imaging system,'' in {\em IEEE Nuclear Science Symposuim
  Medical Imaging Conference}, pp.~1640--1646, Oct 2010.

\bibitem{Wahl2015}
C.~G. Wahl, W.~R. Kaye, W.~Wang, F.~Zhang, J.~M. Jaworski, A.~King, Y.~A.
  Boucher, and Z.~He, ``The polaris-h imaging spectrometer,'' {\em Nuclear
  Instruments and Methods in Physics Research Section A: Accelerators,
  Spectrometers, Detectors and Associated Equipment}, vol.~784, pp.~377 -- 381,
  2015.
\newblock Symposium on Radiation Measurements and Applications 2014 (SORMA XV).

\bibitem{Legge1968}
G.~Legge and P.~V. der Merwe, ``A double scatter neutron spectrometer,'' {\em
  Nuclear Instruments and Methods}, vol.~63, no.~2, pp.~157 -- 165, 1968.

\bibitem{Barnowski2015}
R.~Barnowski, A.~Haefner, L.~Mihailescu, and K.~Vetter, ``Scene data fusion:
  Real-time standoff volumetric gamma-ray imaging,'' {\em Nuclear Instruments
  and Methods in Physics Research Section A: Accelerators, Spectrometers,
  Detectors and Associated Equipment}, vol.~800, pp.~65 -- 69, 2015.

\bibitem{McKisson1994}
J.~E. McKisson, P.~S. Haskins, G.~W. Phillips, S.~E. King, R.~A. August, R.~B.
  Piercey, and R.~C. Mania, ``Demonstration of three-dimensional imaging with a
  germanium compton camera,'' {\em IEEE Transactions on Nuclear Science},
  vol.~41, pp.~1221--1224, Aug 1994.

\bibitem{Poitrasson2014}
A.~Poitrasson-Rivière, M.~C. Hamel, J.~K. Polack, M.~Flaska, S.~D. Clarke, and
  S.~A. Pozzi, ``Dual-particle imaging system based on simultaneous detection
  of photon and neutron collision events,'' {\em Nuclear Instruments and
  Methods in Physics Research Section A: Accelerators, Spectrometers, Detectors
  and Associated Equipment}, vol.~760, pp.~40 -- 45, 2014.

\bibitem{Gozani2009}
T.~Gozani, ``Fission signatures for nuclear material detection,'' {\em IEEE
  Transactions on Nuclear Science}, vol.~56, pp.~736--741, June 2009.

\bibitem{Geiger1975}
K.~Geiger and L.~V.~D. Zwan, ``Radioactive neutron source spectra from 9be(α,
  n) cross section data,'' {\em Nuclear Instruments and Methods}, vol.~131,
  no.~2, pp.~315 -- 321, 1975.

\bibitem{Hamel2016}
M.~Hamel, J.~Polack, A.~Poitrasson-Rivière, M.~Flaska, S.~Clarke, S.~Pozzi,
  A.~Tomanin, and P.~Peerani, ``Stochastic image reconstruction for a
  dual-particle imaging system,'' {\em Nuclear Instruments and Methods in
  Physics Research Section A: Accelerators, Spectrometers, Detectors and
  Associated Equipment}, vol.~810, pp.~120 -- 131, 2016.

\bibitem{Andreyev2011}
A.~Andreyev, A.~Sitek, and A.~Celler, ``Fast image reconstruction for compton
  camera using stochastic origin ensemble approach,'' {\em Medical Physics},
  vol.~38, no.~1, 2011.

\bibitem{scikit-learn}
F.~Pedregosa, G.~Varoquaux, A.~Gramfort, V.~Michel, B.~Thirion, O.~Grisel,
  M.~Blondel, P.~Prettenhofer, R.~Weiss, V.~Dubourg, J.~Vanderplas, A.~Passos,
  D.~Cournapeau, M.~Brucher, M.~Perrot, and E.~Duchesnay, ``Scikit-learn:
  Machine learning in {P}ython,'' {\em Journal of Machine Learning Research},
  vol.~12, pp.~2825--2830, 2011.

\bibitem{Goldsmith2014}
J.~E.~M. Goldsmith, M.~D. Gerling, and J.~S. Brennan, ``{MINER} - a mobile
  imager of neutrons for emergency responders,'' in {\em Symposium on Radiation
  Measurements and Applications}, (Ann Arbor, MI), June 9-12 2014.

\bibitem{Cattaneo2016}
P.~Cattaneo, M.~D. Gerone, F.~Gatti, M.~Nishimura, W.~Ootani, M.~Rossella,
  S.~Shirabe, and Y.~Uchiyama, ``Time resolution of time-of-flight detector
  based on multiple scintillation counters readout by sipms,'' {\em Nuclear
  Instruments and Methods in Physics Research Section A: Accelerators,
  Spectrometers, Detectors and Associated Equipment}, vol.~828, pp.~191 -- 200,
  2016.

\bibitem{Pozzi2012}
S.~Pozzi, S.~Clarke, W.~Walsh, E.~Miller, J.~Dolan, M.~Flaska, B.~Wieger,
  A.~Enqvist, E.~Padovani, J.~Mattingly, D.~Chichester, and P.~Peerani,
  ``Mcnpx-polimi for nuclear nonproliferation applications,'' {\em Nuclear
  Instruments and Methods in Physics Research Section A: Accelerators,
  Spectrometers, Detectors and Associated Equipment}, vol.~694, pp.~119 -- 125,
  2012.

\bibitem{Lebigot2010}
E.~O. Lebigot, ``{Uncertainties}: a {Python} package for calculations with
  uncertainties.''
\newblock Version 2.4.8.1.

\bibitem{Enqvist2013}
A.~Enqvist, C.~C. Lawrence, B.~M. Wieger, S.~A. Pozzi, and T.~N. Massey,
  ``Neutron light output response and resolution functions in ej-309 liquid
  scintillation detectors,'' {\em Nuclear Instruments and Methods in Physics
  Research Section A: Accelerators, Spectrometers, Detectors and Associated
  Equipment}, vol.~715, pp.~79 -- 86, 2013.

\bibitem{Pino2014}
F.~Pino, L.~Stevanato, D.~Cester, G.~Nebbia, L.~Sajo-Bohus, and G.~Viesti,
  ``The light output and the detection efficiency of the liquid scintillator
  ej-309,'' {\em Applied Radiation and Isotopes}, vol.~89, pp.~79 -- 84, 2014.

\end{thebibliography}

\end{document}